\documentclass{aastex}
\usepackage{emulateapj5}
\usepackage{graphics}
\usepackage{epsfig}

\def\preprint{1}    
\def\setfig#1#2{\def#1{#2}
\if\preprint1 #1\fi}
\def\printfig#1{
\if\preprint0
\begin{figure}
#1
\end{figure}
\fi
}
\def\today{\ifcase\month\or
  January\or February\or March\or April\or May\or June\or
  July\or August\or September\or October\or November\or December\fi
  \space\number\day, \number\year}

\slugcomment{Astrophys. J. Lett., revised, \today}

\def\m{\, {\rm m}}
\def\s{\, {\rm s}}

\def\arcsec{{\hbox{\rlap{\rlap{\tt\char"0D}\hbox{\thinspace\tt\char"0D}}
\kern-4.8pt\raise1pt\hbox{$\mit\mathchar"017F$}}}}
\def\CA{{\cal A}}
\def\CP{{\cal P}}

\def\note #1]{{\bf #1]}}
\def\figdir{.}

\begin{document}

\title{Solar-like oscillations of semiregular variables}

\author{J. Christensen-Dalsgaard$^{1,2}$, H. Kjeldsen$^2$, J. A. Mattei$^3$ }
\affil{$^1$Institut for Fysik og Astronomi, Aarhus Universitet, 
DK-8000 Aarhus C, Denmark; jcd@ifa.au.dk}
\affil{$^2$Teoretisk Astrofysik Center, Danmarks Grundforskningsfond;
hans@ifa.au.dk}
\affil{$^3$American Association of Variable Star Observers,
25 Birch St., Cambridge, MA 02138, USA; jmattei@aavso.org}

\begin{abstract}
Oscillations of the Sun and solar-like stars are believed to be
excited stochastically by convection near the stellar surface.
Theoretical modeling predicts that the resulting amplitude
increases rapidly with the luminosity of the star.
Thus one might expect oscillations of substantial amplitudes
in red giants with high luminosities and vigorous convection.
Here we present evidence that such oscillations may in fact
have been detected in the so-called semiregular variables,
extensive observations of which have been made by 
amateur astronomers in the American Association for
Variable Star Observers (AAVSO).
This may offer a new opportunity for studying the physical
processes that give rise to the oscillations, possibly
leading to further information about the properties of convection
in these stars.
\end{abstract}

\keywords{convection -- supergiants -- stars: variables: other --
stars: oscillations}

\section{Introduction}

There is strong evidence that the observed solar oscillations
are intrinsically stable but excited stochastically by the
turbulent convection in the near-surface layers.
Stability calculations that take the interaction with convection into
account generally find linear stability (e.g. Balmforth 1992a);
also, it would be difficult to account for the observed very low
amplitudes if the modes had been unstable (Kumar \& Goldreich 1989).
On the other hand,
calculations based on a mixing-length formulation (Balmforth 1992b)
or detailed hydrodynamical simulations of convection
(e.g. Stein \& Nordlund 1998, 2001) have obtained energy input from
stochastic convective excitation that appears consistent with the
observed oscillation amplitudes and damping rates.
Also, the observed distribution of mode amplitudes 
(e.g. Chaplin et al. 1997) is essentially consistent with
theoretical expectations assuming this excitation mechanism
(Chang \& Gough 1998).

On this basis one may try to predict amplitudes of similar oscillations
in other stars.
Early simple estimates by Christensen-Dalsgaard \& Frandsen (1983) 
indicated that the amplitudes increase with effective temperature
along the main sequence, and also increase rapidly with
increasing luminosity.
These estimates have recently been confirmed by more detailed
calculations, although still based on a mixing-length formalism,
by Houdek et al. (1999).
In the present context it is particularly noteworthy that 
Christensen-Dalsgaard \& Frandsen predicted quite substantial
amplitudes for stars on the red-giant branch.

The predicted amplitudes for main-sequence stars are very low;
a few parts per million in intensity or of order $1 \m \s^{-1}$
in radial velocity.
Thus it is hardly surprising that little evidence is available
for solar-like oscillations in stars near the main sequence.
Brown et al.\ (1991) found indications of solar-like power in Procyon.
This was recently confirmed by Marti\`c et al. (1999) 
and Barban et al.\ (1999) who, however, noted that the inferred
amplitude was smaller by about a factor of 3 than the
predictions of Houdek et al.\ (1999).
Kjeldsen et al. (1995) detected oscillations in the equivalent widths
of Balmer lines in the subgiant $\eta$ Boo, at amplitudes which
were not inconsistent with the theoretical expectations;
however, Brown et al. (1997) were unable to confirm these oscillations
in Doppler-velocity observations.
Gilliland et al.\ (1993) made an extensive search for 
solar-like oscillations in stars in the open cluster M67,
but with no positive detections;
in several cases they obtained upper limits to the amplitudes
substantially below the theoretical predictions.
For the solar near-twin $\alpha$ Cen A Kjeldsen et al. (1999)
failed to find definite oscillations, although the upper limit
was again consistent with expectations.
However, Schou \& Buzasi (2001) found clear evidence for oscillations
in photometric observations 
of $\alpha$ Cen A from the WIRE satellite, with amplitudes
slightly higher than solar, as expected.
Finally, Bedding et al.\ (2001) very recently made a definite detection 
of solar-like oscillations in the sub-giant $\beta$ Hyi,
while Bouchy \& Carrier (2001) obtained a strikingly clear signal
of oscillations in $\alpha$ Cen A,
in both cases with amplitudes close to the expected.
Evidently, the observational situation for stars
near the main sequence is improving, while the
theoretical understanding is still rather uncertain.

For giant stars, larger amplitudes are expected.
Indeed, Arcturus (of spectral type K1.5III)
has been found to show oscillations that are plausibly of solar-like nature 
(Smith, McMillan \& Merline 1987; Innis et al.\ 1988; Merline 1998).
Similar oscillations have also been found in
luminosity observations of K giants in 47 Tuc (Edmonds \& Gilliland 1996),
and in radial-velocity observations of a sample of K and M
giants and supergiants (Cummings et al.\ 1998).
Finally, Buzasi et al.\ (2000) found possible evidence for
solar-like oscillations in the moderate red giant $\alpha$~UMa.

For even more luminous red giants quite substantial amplitudes,
possibly reaching several tenths of a magnitude, are expected.
Studies of these stars are complicated, however, by the fact that
periods of the order of several months or even years are expected,
requiring very extended observations which are 
generally not available from professional investigations.
It is therefore extremely fortunate that the American Association of
Variable Star Observers (AAVSO) has a very extensive database
on large-amplitude, long-period pulsators.
In this letter, we analyze a sample of such stars to argue that
they do indeed show evidence for stochastically excited, solar-like
oscillations.

\section{Observations of semiregular variables}

The observations were obtained from the AAVSO archives,
in most cases extending over 30 -- 40 years;
although more extensive data are available, these are so far the longest
sequences that have been validated and analyzed in detail.
The CLEANest algorithm (Foster 1995) was used, involving least-squares fitting
of the data to sinusoids taking into account the possible
presence of multiple periods.
The data were analyzed over windows of typical lengths
of a few thousand days,
hence providing measures of the periods $P(t)$ and
amplitudes $A(t)$ as functions of time $t$.
From this we can in particular determine the mean 
$\langle A \rangle$ and the root-mean-square standard
deviation $\sigma(A)$ of the amplitudes.

Some results are illustrated in Fig.~\ref{fig:obs1}.
As noted already by Mattei et al. (1997) there is a clear
division in this diagram, between the Mira variables
with large amplitudes and small amplitude scatter
and the semiregular (SR) variables with relatively
low amplitudes and large scatter.
Also, it is striking that there appears to be a strong correlation,
in the latter group, between $A$ and $\sigma(A)$, whereas no
such correlation is found for the Miras.
This correlation is the main piece of evidence which we 
consider in the following.

\setfig{\obsone}{
{\begin{center}
\leavevmode
\centerline{\epsfxsize=8.0cm \epsfbox{\figdir/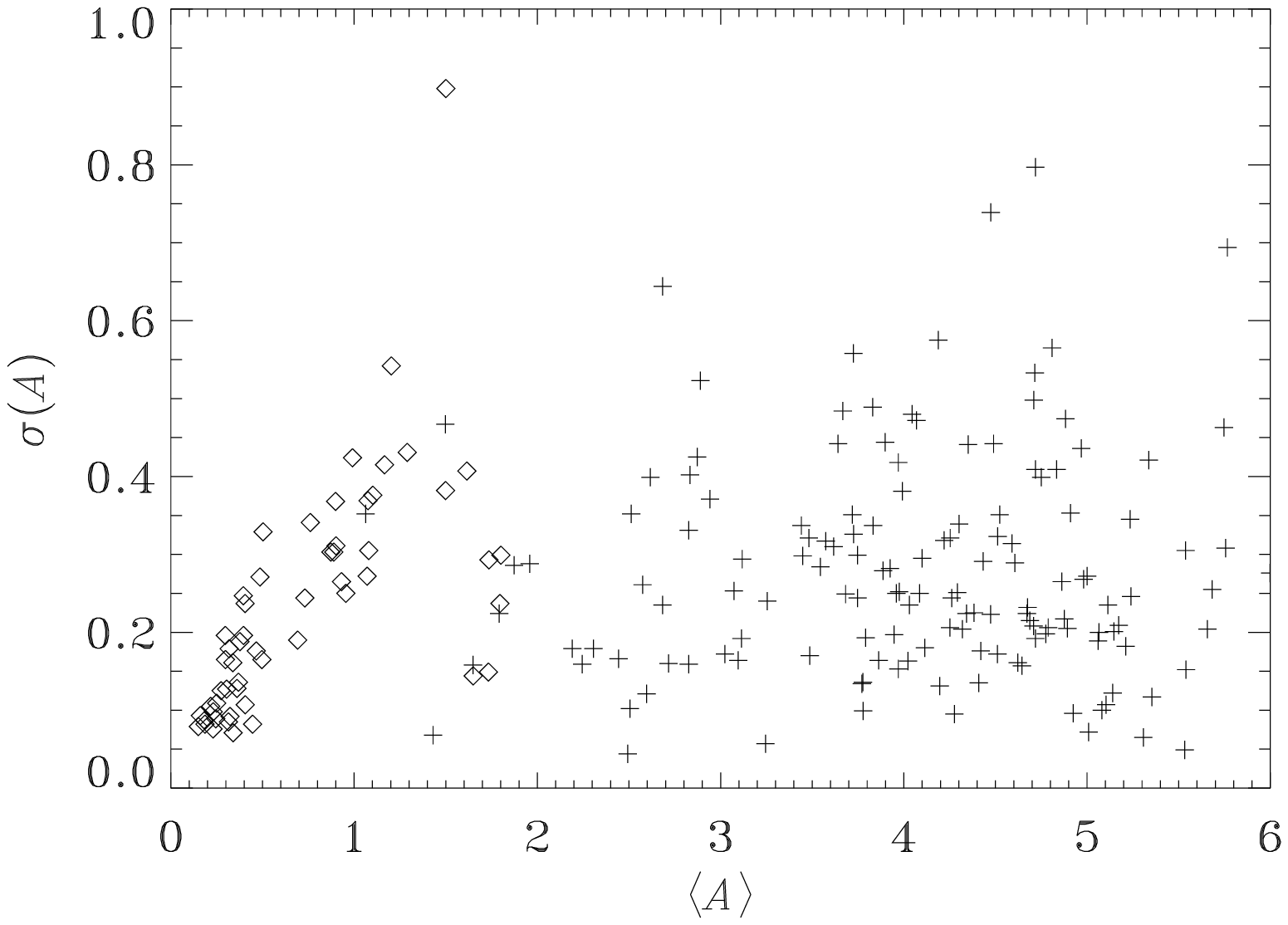}}
\end{center}
\figcaption{
Properties of long-period variables, from AAVSO observations
analyzed with the CLEANest algorithm.
The abscissa is amplitude $A$ in visual magnitude, and the ordinate
is the root-mean-square scatter in the amplitude.
Crosses and diamonds show stars that have been classified
as Mira variables and semiregular variables, respectively.
Adapted from Mattei et al.\ (1997).
\label{fig:obs1}
}
}
}

\section{Properties of stochastically excited oscillators}

%
The statistical properties of stochastically excited oscillations
were discussed, for example, by Kumar, Franklin \& Goldreich (1988)
and Chang \& Gough (1998).
If the modes are observed over a time short compared with
the damping time, the energy distribution is exponential,
leading to the following distribution for the amplitude $A$:
\begin{equation}
p(A) = {2 A \over \CA^2} \exp\left( - {A^2 \over \CA^2}\right) \; ,
\label{eqn:dist}
\end{equation}
where the average pulsation energy is proportional to $\CA^2$.
The observed amplitudes of solar oscillation satisfy this
distribution quite accurately
(e.g. Chaplin et al. 1997).
According to Eq. (\ref{eqn:dist})
the average $\langle A \rangle$ and standard 
deviation $\sigma(A)$ of the amplitude are determined by
$\langle A \rangle = (\sqrt{\pi} / 2) \CA$,
$\sigma^2(A) =  \langle A^2 \rangle  - \langle A \rangle^2 
= (1 - \pi / 4 ) \CA^2$,
and hence
\begin{equation}
\sigma(A) =  \left({4 \over \pi} - 1 \right)^{1/2} \langle A \rangle 
\simeq 0.52 \langle A \rangle  \; .
\label{eqn:std}
\end{equation}

%
To illustrate this behavior, we have computed a set of artificial
data simulating a stochastically excited oscillator with a period of 82 days.
The simulation is in addition characterized by the sampling-time interval
$\Delta t$, the damping time $\tau_{\rm d}$ and the average 
excitation $\delta A$ at each sampling time.
The signal is written as 
$A_1(t) \sin(\omega t) + A_2(t) \cos(\omega t)$,
where $\omega$ is the angular frequency corresponding to the period.
At each sampling time the amplitudes $A_1$ and $A_2$ are reset,
according to
\begin{equation}
A_i(t+\Delta t) = \exp( - \Delta t/\tau_{\rm d}) A_i(t)
+ \delta A \, {\cal N}_i \; ,
\label{eqn:excit}
\end{equation}
where ${\cal N}_i$ is a normally distributed random variable
with zero mean and unit variance.
The `data' cover a total period of around 1600 years.
They were analyzed in segments of 1 year, to determine the total
power $\CP_i$ for each segment which was used as an estimate
of the mode energy; 
the energy, normalized to its average, was then determined as
$E_i / \langle E \rangle = \CP_i / \langle \CP \rangle$,
where the average is over the segments.
The mean and standard deviation of the amplitudes,
estimated by $\sqrt{\CP_i}$, were similarly evaluated.

Figure~\ref{fig:art} shows an example of a simulated time\-string,
together with the distribution of inferred mode energies, binned
and suitably normalized (cf. Chang \& Gough 1998)
to make it directly comparable with
the exponential distribution $\exp( - E / \langle E \rangle)$ which is
also shown.
It is evident that the data do indeed approximately satisfy the
exponential distribution of energy.
For this sample, we find $\sigma(A) / \langle A \rangle = 0.49$,
in reasonable agreement with Eq.~(\ref{eqn:std}).

\section{Results and discussion}

%
According to Eq.~(\ref{eqn:std}) there is a direct relation between
the value and scatter of the amplitude for stochastically excited
oscillators.
This is clearly reminiscent of the correlation observed in
Fig.~\ref{fig:obs1} amongst the SR variables.
To make a more detailed comparison, we show results for these
variables again in Fig.~\ref{fig:obs2}, together with the 
relation (\ref{eqn:std}).
It is evident that the observed and predicted behavior are
indeed quite similar.
Thus it is plausible that the observed oscillations are 
excited in this manner.

\setfig{\art}{
{\begin{center}
\centerline{\epsfxsize=8.0cm \epsfbox{\figdir/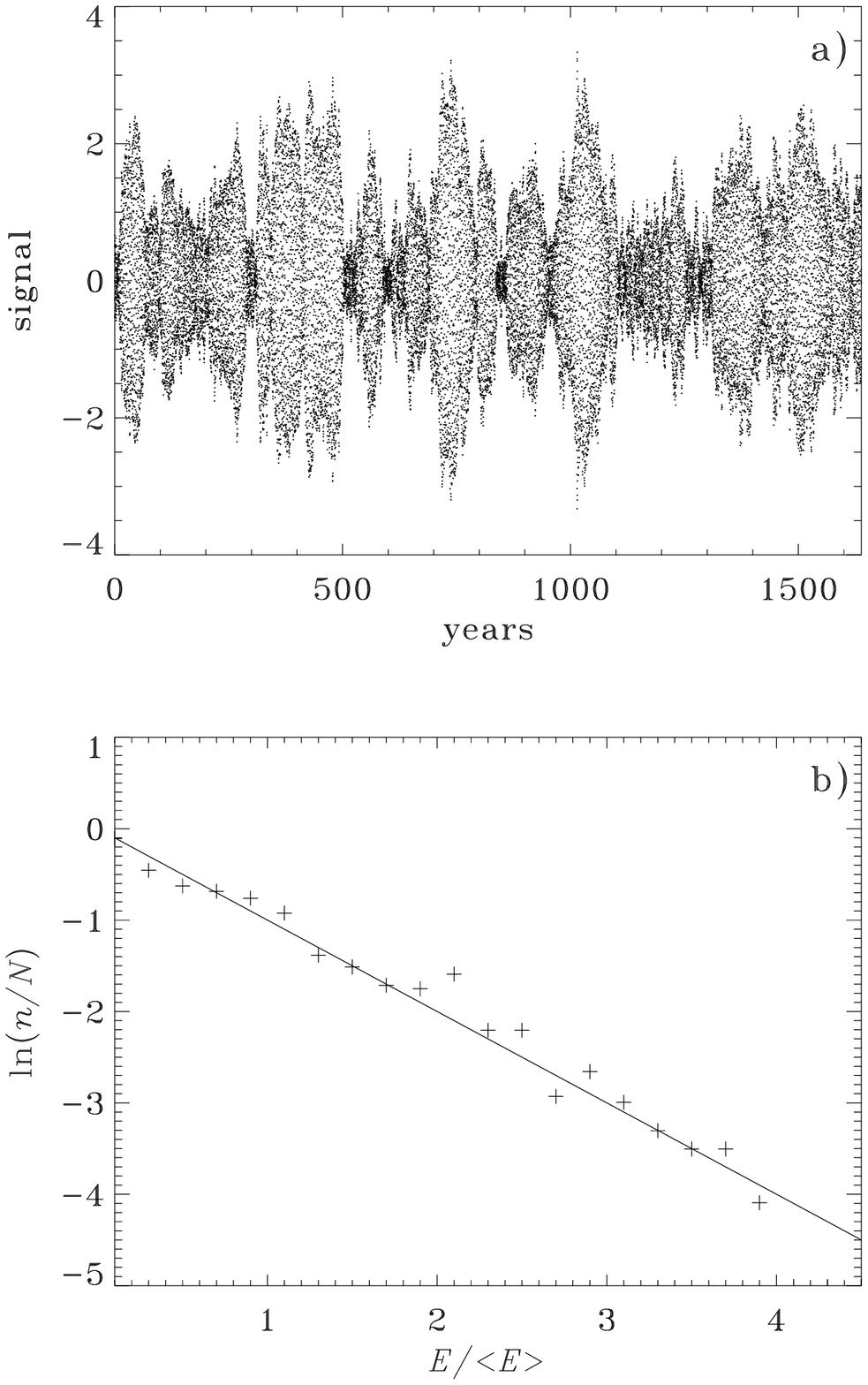}}
\end{center}
\figcaption{
Artificial time series for an oscillation with a period of 82 days,
a damping time $\tau_{\rm d}$ of 60~years
and a sampling-time interval $\Delta t$ of 20 days,
computed according to Eq.~(\ref{eqn:excit}).
The top panel shows the computed time series which, as indicated,
covers about 1600 years.
In the bottom panel the points show the binned, normalized distribution
of mode energy, in units of the mean energy; the line corresponds to the
expected exponential distribution in Eq. (\ref{eqn:dist})
(for further details, see Chang \& Gough 1998).
\label{fig:art}
}
}
\bigskip
}

In the case of the Sun, and the carefully observed K giant Arcturus,
a substantial number of modes are observed to be excited,
as would indeed be expected
given the broad-band nature of the stochastic forcing.
Christensen-Dalsgaard \& Frandsen (1983) noted that
the number of available modes decreases in very luminous giant
stars where the acoustical cut-off frequency, which likely 
provides an upper limit to the observable spectrum, approaches
the frequency of the fundamental radial mode.
Even so, it is encouraging that most
of the SR variables appear to be at least double-periodic
(Mattei et al. 1997).

Further analyses are required to reach 
a definite conclusion on the nature of these stars.
Additional scatter in the inferred amplitudes might arise
from beating between unresolved frequencies in the 
spectra of the stars.
On the other hand, it was noted by Kumar et al. (1988) that 
the amplitude distribution becomes more peaked, reducing the scatter,
if the observing time is not short compared with the 
(in this case entirely unknown) lifetime of the modes.
An additional effect may arise because of the large
amplitudes of the oscillations.
Indeed, it is likely that for the SR variables, as for the Mira
variables, the amplitude in the visible region of the spectrum
is strongly enhanced by atmospheric effects.
By comparing the AAVSO amplitudes with the bolometric
amplitudes obtained from Whitelock, Marang \& Feast (2000),
for the stars common between the two samples, we find that
the visual amplitudes are on average larger by a factor 5 -- 6 than
the bolometric amplitudes.
This may cause a nonlinear
relation between the underlying pulsations and the observed response,
possibly involving saturation at the highest amplitudes.
To model this we have carried out additional analyses of
the simulated timestring shown in Fig.~\ref{fig:art}a, after
filtering with a suitably scaled ${\rm sinh}^{-1}$ response.
It was found that such saturation reduced the scatter quite substantially:
even if the maximum amplitude was reduced by only 17 per~cent,
relative to the linear response, $\sigma(A)/\langle A \rangle$
was reduced from the value of 0.49 found in the linear case to 0.47.
There is indeed some hint that the observed values are below 
the linear trend at higher amplitudes, in accordance with this mechanism.

\setfig{\obstwo}{
\begin{center}
\centerline{\epsfxsize=8.0cm \epsfbox{\figdir/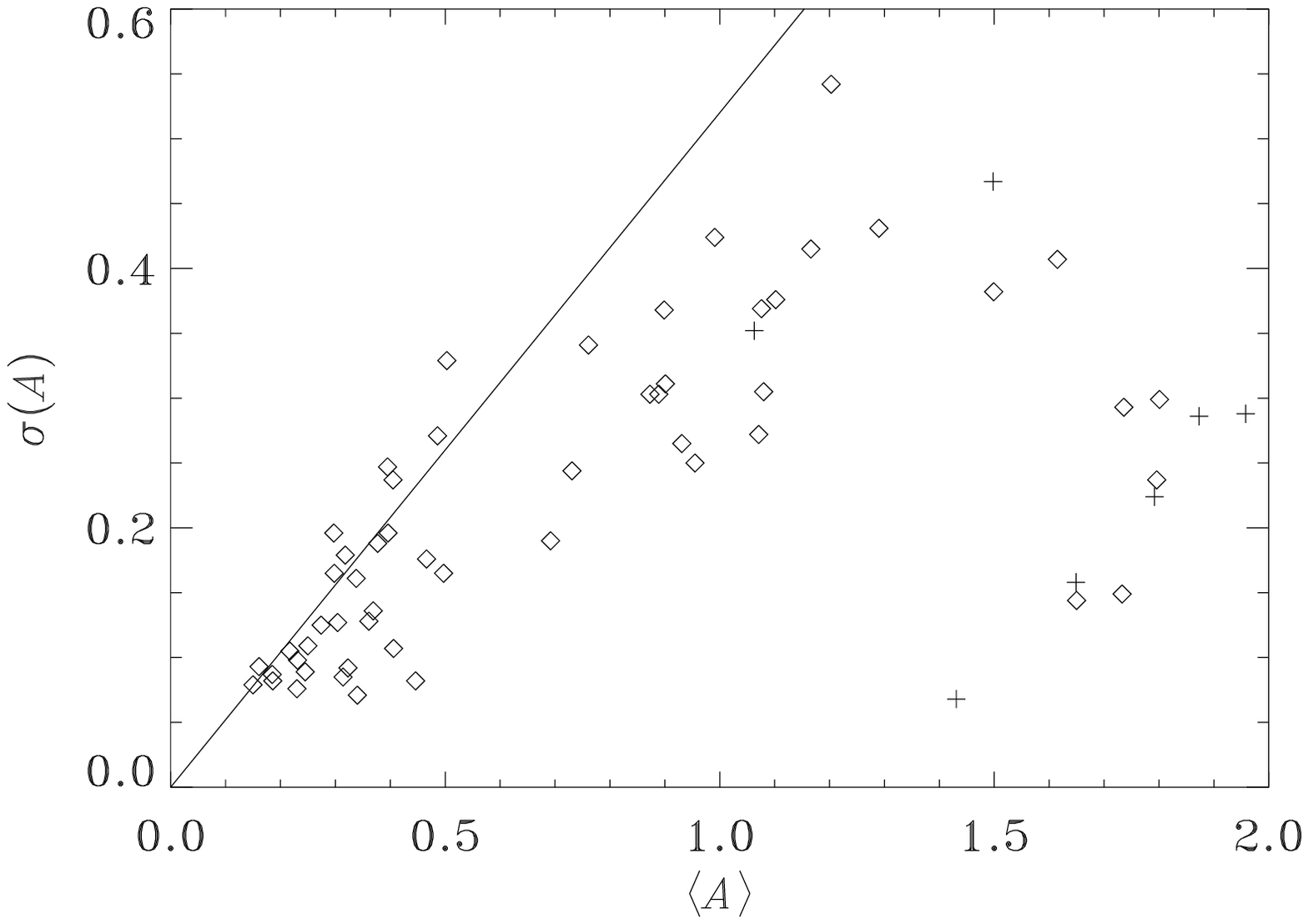}}
\end{center}
\figcaption{
Root-mean-square scatter in the amplitude against amplitude, for
the low-amplitude sample of AAVSO stars (see also Fig.~\ref{fig:obs1}).
Crosses and diamonds show stars that have been classified
as Mira variables and semiregular variables, respectively.
The solid line shows the theoretical relation for stochastically
excited oscillators, from Eq.~(\ref{eqn:std}).
\label{fig:obs2}
}
\bigskip
}

The amplitudes found for the semiregular variables range as high
as around 1.5 magnitude in the visual (cf.\ Fig.~\ref{fig:obs2}).
Assuming the scaling between visual and bolometric amplitudes
mentioned above, this corresponds to an amplitude in relative
luminosity variations $\delta L/L$ of around 0.3,
roughly five orders of magnitude higher than observed in the Sun.
No estimates have so far been made of amplitudes of specific
stars in this group. 
Based on the calculations by Christensen-Dalsgaard \& Frandsen (1983),
Kjeldsen \& Bedding (1995) estimated
that $\delta L/L$ scales as $L M^{-1} T_{\rm eff}^{-1}$, 
where $L$, $M$ and $T_{\rm eff}$ are luminosity, mass and effective
temperature of the star.
The validity of extrapolating this relation to the parameter
range of the semiregular variables is perhaps somewhat doubtful;
also, the parameters of these stars, particularly their mass,
are uncertain.
However, the amplitudes observed are at least not obviously
inconsistent with the estimates based on the scaling relation.

\vfill\eject

\section{Conclusions}

At present we have secure observational information about very few
stochastically excited pulsating stars,
and hence additional information about this kind of pulsation,
for a broad range of stars, would be extremely valuable,
as a test for our models of the excitation.
With a better physical understanding of the excitation mechanism
the properties of the oscillations may also provide interesting
diagnostics of the properties of convection, as a supplement
to direct observations of velocity and intensity effects of
surface convection.
Secure identification of stochastically excited pulsations
in the semiregular variables would therefore
evidently be of substantial interest.

The present study is of a statistical nature.
To confirm our suggestion a more detailed analysis of individual
stars, including careful modeling of the stellar structure and
the properties of the oscillations, will be required,
furthermore using the frequencies to constrain the overall 
properties of the stars.
Also, further observational data should be sought,
making full use of the AAVSO data archive, and
of other similar archives of long-term observations.
Indeed, the century-long data series available for a number of stars
in the AAVSO archive are now being validated and quality controlled;
these full datasets will then be submitted to the same 
type of analysis as reported here.
To obtain additional observations of these long-period
variables will clearly be a long-term project, although
one definitely worth undertaking.
However, it is also of evident interest to study similar
variations in stars further down the red-giant branch,
such as already observed in K giants, of shorter periods.
Here the predicted amplitudes are also much smaller,
precluding visual observation of the variations.
However, with the present trend towards increasing sophistication
of amateur equipment, including the extensive use of CCD detectors,
there is little doubt that amateur astronomers will be able to
continue and extend their very valuable efforts towards providing 
long-term observations of a nature that are difficult to obtain in
a professional setting.

\section*{Acknowledgements}

We are grateful to the numerous observers whose dedicated work
forms the basis for the present study.
F. P. Pijpers is thanked for useful conversations.
We thank C. \.{I}bano\u{g}lu for organizing the 1998 NATO Advanced Study
Institute in {\c C}e{\c s}me, Turkey, where the original idea for this
work originated.
This work was supported in part by the Danish National Research
Foundation through its establishment of the Theoretical Astrophysics Center.

\if\preprint0
\vfill\eject
\printfig{\obsone}
\printfig{\art}
\printfig{\obstwo}
\fi

\end{document}